\documentclass[aps,pra,a4paper,amsmath,amssymb,tightenlines,superscriptaddress,twocolumn,showpacs]{revtex4-1}
\usepackage{graphicx}
\usepackage{bm} 
\usepackage{dcolumn}
\usepackage{bm}
\usepackage{color,ulem}
\usepackage{enumerate}
\usepackage[T1]{fontenc}

\newcommand{\bk}{\mathbf{k}}
\renewcommand{\r}{\mathbf{r}}
\newcommand{\tnc}{n^{\rm{c}}}
\newcommand{\tnnc}{n^{\rm{nc}}}

\def\nnc{{\bm n}^{\rm nc}}
\def\nc{{\bm n}^{\rm c}}
\def\tmu{\tilde{\mu}}
\def\tp{\tilde{p}}
\def\dc{d^{\rm c}}
\def\dnc{d^{\rm nc}}

\begin{document}
\title{Finite temperature phase diagram of a spin-1 Bose gas}
\author{Yuki Kawaguchi}
\author{Nguyen Thanh Phuc}
\affiliation{Department of Physics, University of Tokyo, 7-3-1 Hongo, Bunkyo-ku, Tokyo 113-0033, Japan}
\author{P.~Blair Blakie}
\affiliation{Jack Dodd Centre for Quantum Technology, Department of Physics, University of Otago, Dunedin, New Zealand}

\date{\today}
\begin{abstract}  
We formulate a self-consistent Hartree-Fock theory for a spin-1 Bose gas at finite temperature and apply it to characterizing the phase diagram. 
We find that spin coherence between thermal atoms in different magnetic sub-levels develops via coherent collisions with the condensed atoms, and is a crucial factor in determining the phase diagram.
We develop analytical expressions to characterize the interaction and temperature dependent shifts of the phase boundaries.
\end{abstract}


\maketitle

\section{Introduction}
\begin{figure}[!tbh]
\begin{center}
\includegraphics[width=3.3in]{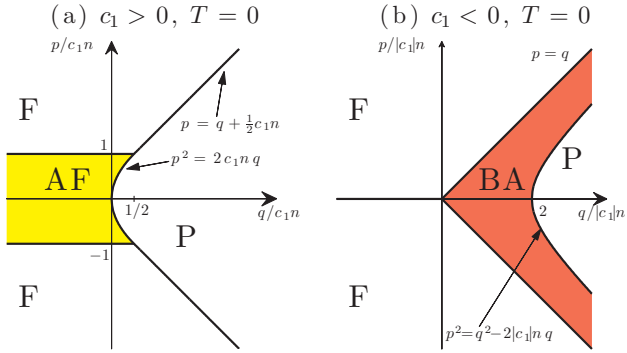} 
\caption{\label{phasediag}
(Color online)
 The $T=0$ phase diagram of a spin-1 Bose gas for cases where the spin dependent interaction is (a)  {antiferromagnetic} ($c_1>0$) and (b)  {ferromagnetic} ($c_1<0$). {The   vertical  and horizontal axes are the linear and quadratic Zeeman energies } (see text) in units of   $|c_1|n$, where $n$ is the total number density (which is identical to the  condensate number density  at $T=0$).  The phases shown are (F) ferromagnetic, (P) polar, (AF) antiferromagnetic, and (BA) broken-axisymmetry phases  (see Sec.~\ref{phaseID} and Refs.~\cite{Stenger1999a,Ueda2010R}).
The rotational symmetry about the direction of the applied field is spontaneously broken in the AF and BA phases.
}
\end{center}
\end{figure}

A key feature of a system with spin internal degrees of freedom is that the atoms can condense into a range of phases, characterized by various spin order parameters, dependent upon the nature of the interactions and the external magnetic field (e.g.~see Fig.~\ref{phasediag}). The seminal theory for the spin-1 Bose gas was developed in 1998 \cite{Ho1998a,Ohmi1998a} and soon after realized in experiments \cite{Stenger1999a,Miesner1999a}. 
Aspects of the equilibrium phase diagram were initially observed in Ref.~\cite{Stenger1999a}, and more recently experiments have used external fields to investigate the dynamical properties of this system (e.g.~see Refs.~\cite{Chang2004a,*Chang2005a,Black2007a,Vengalattore2008a,*Vengalattore2010a,Liu2009b}), including quenches between phases \cite{Sadler2006a,Liu2009a}.

Several theoretical treatments within mean-field approximations have considered the equilibrium properties of a condensed spin-1 Bose gas at finite temperature \cite{Isoshima2000a,Huang2002a,Zhang2004a,KisSzab2007a,Phuc2011a}.  Natu and Mueller have predicted that, for sufficiently large spin dependent interaction strength, pairing or spontaneous magnetization will occur at slightly higher temperature than the condensation transition~\cite{Natu2011a}. In the 2D regime, where condensation is expected to be suppressed, the finite temperature phase diagram has recently been elucidated \cite{Mukerjee2006a,James2011a}.

This paper investigates the finite-temperature phase diagram of the spin-1 Bose gas,
including both linear and quadratic Zeeman effects,
which were not fully considered in previous work~\cite{Isoshima2000a,Huang2002a,Zhang2004a,Phuc2011a}.
Figure~\ref{phasediag} shows the mean-field phase diagram at $T=0$ drawn in the parameter space of the linear ($p$) and quadratic ($q$) Zeeman energies~\cite{Stenger1999a,Ueda2010R}.
We investigate how the phase boundaries in Fig.~\ref{phasediag} change as temperature increases,
using a Hartree-Fock (HF) mean-field theory.
Although HF theory is the simplest many-body theory, it forms an important building block for more advanced  many-body  theories, and for comparison to other types of calculations.
A key feature of our theory is the inclusion of spin coherence between non-condensate (thermal atoms) in different magnetic sub-levels. We find that when the condensate is in a state of spontaneously broken spin rotational symmetry (about the direction of the applied field), i.e., in the antiferromagnetic (AF) and broken-axisymmetry (BA) phases, the spin coherence between non-condensed atoms also develops via coherent collisions with the condensed atoms.
Moreover, the non-condensate spin coherence has a large effect on the phase boundaries in the finite temperature regime.
We derive analytic relations between the shifts in the phase boundaries and non-condensate spin density or spin coherence, which agree well with the full numerical results.
These analytic results furnish additional insight into how the thermal fluctuations influence the condensate order and directly show the importance of the non-condensate spin coherence.

Finally, we note that HF calculations are generally expected to provide a good qualitative description of the interacting system. Indeed,  HF theory accurately describes a range of  thermodynamic measurements made on the scalar three-dimensional Bose gas (e.g.~see \cite{Gerbier2004a,Gerbier2004b,Tammuz2011a}).
However, the spinor situation is much less clear. Our recent work  \cite{Phuc2011a} suggests that the spinor gas, in the  regime of current experiments with $^{87}$Rb, is strongly interacting, in the sense that the corrections to the Bogoliubov theory are non-perturbative.  There also remain a number of open questions about the explanation of current experiments (e.g.~see \cite{Liu2009a,Vengalattore2010a}) and what role thermal fluctuations, dipole-dipole interactions, or non-equilibrium effects play. The work we present here provides an important step towards achieving a more complete understanding of thermal effects in the spinor Bose gas.

\section{Basic formalism} 
We consider a spin-1 Bose gas confined in an optical potential $U(\mathbf{r})$ and subject to a uniform magnetic field along $z$. The single particle description of the atoms is provided by the Hamiltonian
\begin{equation}
(h_0)_{ij}=\left[-\frac{\hbar^2\nabla^2}{2M}+U(\mathbf{r})-pi+qi^2\right]\delta_{ij},
\end{equation}
where $p$ and $q$ are the coefficients of the linear and quadratic Zeeman terms, respectively,  the subscripts $i,j=-1,0,+1$, refer to the magnetic sub-levels of the atoms, and  $M$ is the atomic mass.
The value of $q$ is tunable independently of $p$,
using an off-resonant microwave field~\cite{Gerbier2006}.

Introducing spinor field operators  $\hat\psi_i(\mathbf{r})$   the  cold-atom Hamiltonian, including interactions, is given by \cite{Ho1998a,Ohmi1998a}
\begin{align}
\hat{\mathcal{H}}\!=\!&\int \!d\mathbf{r}\!\left\{\!\sum_{i,j}\left[\hat\psi_i^\dagger(\mathbf{r})(h_0)_{ij}\hat\psi_j(\mathbf{r})+\frac{c_0}{2}\hat\psi_i^\dagger(\mathbf{r})\hat\psi_j^\dagger(\mathbf{r})\hat\psi_j(\mathbf{r})\hat\psi_i(\mathbf{r})\right]\right. \nonumber\\
&\left.+\frac{c_1}{2}\sum_{\alpha,i,j,k,l}(f_{\alpha})_{ij}(f_{\alpha})_{kl} \hat\psi_i^\dagger(\mathbf{r})\hat\psi_k^\dagger(\mathbf{r})\hat\psi_l(\mathbf{r})\hat\psi_j(\mathbf{r})\right\}
\label{eq:H_org}
\end{align}
where $\alpha=x,y,$ or $z$ specifies the spin components, with $f_\alpha$ being the {$3\times3$} spin-1 matrices.   The  parameters $c_0$ and $c_1$ are referred to as the spin independent and spin dependent interaction parameters, respectively, and are given by $c_0=4\pi\hbar^2(a_0+2a_2)/3M$, $c_1=4\pi\hbar^2(a_2-a_0)/3M$, with $a_S$ ($S=0,2$) being the $s$-wave scattering length for the scattering channel of total spin $S$.

\section{Hartree-Fock theory} 
\subsection{General inhomogeneous theory}
The basic mean-field approach is to assume that when there is a condensate in the system the field operator can be decomposed as
 \begin{equation}
\hat\psi_i(\mathbf{r})=\phi_i(\mathbf{r})+\hat{\delta}_i(\mathbf{r}),
 \end{equation}
 where $\phi_i(\mathbf{r})$ is a classical field describing the condensate and the fluctuation operator, $\hat{\delta}_i(\mathbf{r})$, describes the non-condensate modes.  
The HF equations can be derived by using a variational approach to minimize the free energy (e.g.~see  Appendix \ref{AppdHF}  and Refs.~\cite{QTFS,Bergeman1997a}). Key to this approach is the factorization of the expectation value of the interaction terms into expressions involving products of first order correlation functions \begin{equation}
\langle \hat\psi_i^\dagger(\mathbf{r})\hat\psi_j(\mathbf{r})\rangle=\tnc_{ij}(\mathbf{r})+\tnnc_{ij}(\mathbf{r}),
\end{equation}
where we have introduced the notation $\tnc_{ij}(\mathbf{r})\equiv\phi_i^*(\mathbf{r})\phi_j(\mathbf{r})$ and $\tnnc_{ij}(\mathbf{r})\equiv\langle\hat{\delta}_i^\dagger(\mathbf{r})\hat{\delta}_j(\mathbf{r})\rangle$ for the condensate and non-condensate {one-body} density matrices, respectively~\footnote{The full one-body density matrix also retains off-diagonal position arguments, however these are not needed to formulate HF theory for a gas with contact interactions. In this work we will use the term \textit{off-diagonal} in reference  to the spin indices.}.
We emphasize that in the presence of a condensate, $\tnnc_{ij}(\mathbf{r})$ may have nonzero off-diagonal elements, that is, exhibits partial phase coherence between thermal atoms in different magnetic sub-levels.
Since Hamiltonian~\eqref{eq:H_org} is invariant under a spin rotation about the $z$ axis,
$\tnnc_{ij}(\mathbf{r})$ should be diagonal in the normal phase (without pairing nor ferromagnetic order~\cite{Natu2011a}) so that the system is invariant under spin rotations.
In a condensed phase, however, if the condensate spontaneously breaks the rotational symmetry in spin space,
the non-condensate also distributes inhomogeneously in spin space
due to coherent collisions between condensed and non-condensed atoms.
The non-condensate spin coherence was experimentally observed in a two-component Bose gas~\cite{McGuirk2003a,*Lewandowski2003a}.

The generalized Gross-Pitaevskii equation (GPE) for the condensate is (see Appendix \ref{AppdHF})
\begin{align}
\mu\phi_i(\mathbf{r})=&\sum_{j}  {L}_{ij}\phi_j(\mathbf{r}),\label{genGPE}
\end{align}
where 
\begin{align}
 {L}_{ij}&= (h_0)_{ij}+c_0(\tnc+\tnnc)\delta_{ij}+c_0\tnnc_{ji}  \nonumber\\
&+c_1\sum_{\alpha}\left[(F^{\rm{c}}_\alpha+F^{\rm{nc}}_\alpha)(f_\alpha)_{ij}
+\sum_{k,l}(f_\alpha)_{ik}(f_\alpha)_{lj}\tnnc_{lk}\right],
\end{align}
is the Gross-Pitaevskii matrix operator, and 
\begin{align}
n^{\rm{c}}(\mathbf{r})&=\sum_i\tnc_{ii}(\mathbf{r}),\\
F_\alpha^{\rm{c}}(\mathbf{r})&= \sum_{i,j} (f_\alpha)_{ij}\tnc_{ij}(\mathbf{r}),\\
n^{\rm{nc}}(\mathbf{r})&=\sum_i\tnnc_{ii}(\mathbf{r}),\\
F_\alpha^{\rm{nc}}(\mathbf{r})&=\sum_{i,j}(f_\alpha)_{ij}\tnnc_{ij}(\mathbf{r}),
\end{align}
are the number and spin densities associated with the condensed and non-condensed atoms.

The HF grand canonical Hamiltonian for the non-condensate is given by (see Appendix \ref{AppdHF})
\begin{equation}
K_{\rm{HF}}=\int d\mathbf{r}\sum_{i,j}\hat{\delta}_i^\dagger A_{ij}(\mathbf{r})\hat{\delta}_{j},\label{KHF}
\end{equation}
where
\begin{equation}
A_{ij} =  {L}_{ij}-\mu\delta_{ij}+c_0\tnc_{ji}
+c_1\sum_{k,l}(f_\alpha)_{ik}(f_\alpha)_{lj}\tnc_{ {lk}},
\label{AHF}
\end{equation}
i.e.,~differing from the condensate operator, $L_{ij}$, by the inclusion of the exchange interactions with the condensate.

By finding the eigenvalues ($\epsilon_{\lambda}$) and eigenvectors [$u_j^{(\lambda)}(\mathbf{r})$] of $A_{ij}$, i.e.,
\begin{equation}
\epsilon_{\lambda}u_i^{(\lambda)}(\mathbf{r})=\sum_{j}A_{ij}(\mathbf{r})u^{(\lambda)}_j(\mathbf{r}),
\label{eigeneqA}
\end{equation}
normalized so that
\begin{equation}
\sum_i\int d\mathbf{r}\,u^{(\nu)*}_i(\mathbf{r})u_i^{(\lambda)}(\mathbf{r})=\delta_{\nu\lambda},
\label{OrthNorm_u}
\end{equation}
the non-condensate density matrix is given by
\begin{equation}
\tnnc_{ij}(\mathbf{r})=\sum_{\lambda}u_i^{(\lambda)*}(\mathbf{r})u^{(\lambda)}_j(\mathbf{r})\bar{n}_\lambda,\label{nc_spindm}
\end{equation}
where $\bar{n}_{\lambda}=1/[\exp(\beta \epsilon_{\lambda})-1]$ is the Bose-Einstein distribution {function} with $\beta=1/(k_{\rm B}T)$.

\subsection{Specialization to the uniform system}
For the purpose of studying the finite temperature phase diagram we now discuss the specialization of the HF formalism to  {a} uniform system.
In this case $U(\r)\to0$ and  the mean-fields  {($\tnc_{ij}$ and $\tnnc_{ij}$)} are spatially independent.  The condensate occurs in the zero-momentum spatial mode, and the generalized GPE (\ref{genGPE}) reduces to the nonlinear algebraic equation
\begin{equation}
\mu\phi_i = \sum_{j}  \mathcal{L}_{ij}\phi_j ,\label{genGPEu}
\end{equation}
where
\begin{align}
\mathcal{L}_{ij}=&(-pi+qi^2)\delta_{ij} +c_0\left[(\tnc+\tnnc)\delta_{ij}+ \tnnc_{ji}\right] \! \label{Liju}\\
+&\!c_1\sum_{\alpha}\!\left[(F^{\rm{c}}_\alpha\!+\!F^{\rm{nc}}_\alpha)(f_\alpha)_{ij}
\!+\!\sum_{k,l}(f_\alpha)_{ik}(f_\alpha)_{lj} \tnnc_{ {lk}} \right].\nonumber
\end{align} 

The excited modes have plane wave spatial dependence:
\begin{equation}
u^{(\lambda)}_j(\r)=\bar{u}_j^{(\nu)}e^{i\bk\cdot\r},\label{uUniform}
\end{equation}
where $\bar{u}_j^{(\nu)}$ is a constant spinor (and is independent of $\mathbf{k}$) and  we have adopted the notation $\lambda\to\{\nu,\bk\}$,
with $\bk$ a wave vector and $\nu$   an index to distinguish between modes.

The HF Hamiltonian takes the form
\begin{equation}
A_{ij}=-\frac{\hbar^2\nabla^2}{2M}+\mathcal{A}_{ij},\label{uHF}
\end{equation}
where
\begin{align}
\mathcal{A}_{ij}&=\mathcal{L}_{ij}-\mu\delta_{ij} +c_0\tnc_{ji}  +\!c_1 \sum_{\alpha,k,l}(f_\alpha)_{ik}(f_\alpha)_{lj} \tnc_{{lk}},\label{Aij}
\end{align}
is a constant matrix.
Notably the spatial and spin parts in Eq.~(\ref{uHF})  {are decoupled} and can be treated separately [allowing us to use the excited mode  {of the} form given in Eq.~(\ref{uUniform})]. Diagonalizing $\mathcal{A}_{ij}$  we obtain the three eigenvectors $\bar u^{(\nu)}_j$   with respective eigenvalues $\kappa_{\nu}$, and hence  that the excitation spectrum is  {given by}
\begin{equation}
\epsilon_{\nu\bk}=\frac{\hbar^2k^2}{2M}+\kappa_{\nu}.
\end{equation}
To evaluate the non-condensate  {one-body} density matrix we set $\sum_{\lambda}\to (2\pi)^{-3}\sum_{\nu}\int d\mathbf{k}$ in Eq.~(\ref{nc_spindm}) and obtain
\begin{equation}
\tnnc_{ij}=\sum_{\nu=1}^3 \bar u^{(\nu)*}_i\bar u^{(\nu)}_j\,\frac{ {{\rm Li}_{3/2}(e^{-\beta\kappa_{\nu}})} }{\lambda_{\rm{dB}}^3},
\label{eq:tnnc_Li}
\end{equation}
where $\lambda_{\rm{dB}}=h/\sqrt{2\pi Mk_BT}$ is the thermal de Broglie wavelength and 
 {${\rm Li}_\sigma(z)\equiv \sum_{t=1}^\infty z^t/t^\sigma$ is the polylogarithm}. We note that for the thermal cloud to saturate, and hence condensation to occur, at least one of the eigenvalues $\kappa_{\nu}$ must approach zero at the condensation temperature.

\section{Results}

The effect of the thermal cloud on the condensate is, in general, quite complicated and requires the full self-consistent calculation. 
{We numerically solve the coupled Gross-Pitaevskii and HF equations self-consistently in the temperature range of} $T=(0-0.5)T_0$, where
$T_0$ is the condensation temperature of an ideal scalar gas with the same total number density.  
Because there are three internal states,
the condensation temperature of an ideal spin-1 gas at 
$p=q=0$ is  {reduced to} $T_{\rm c}^{\rm{spinor}}= (1/3)^{2/3}T_0 \simeq 0.48T_0$.
For $^{87}$Rb and $^{23}$Na gases (in the $F=1$ hyperfine multiplet) the spin dependent interaction is small relative to the spin independent interaction  ($c_0\sim10^2 |c_1|$), however for generality we explore larger values of up to $c_1/c_0=\pm0.5$, which might be realizable with new species of atoms, or using magnetic or optical manipulation of inter-atomic interactions.

\subsection{Identification of phases}\label{phaseID}
For the system we consider here of a spin-1 Bose gas subject to a magnetic field,
a variety of phases arise and are well-characterized for the $T=0$ case (see Fig.~\ref{phasediag}).
These phases are identified according to the properties of the condensate order parameter $(\phi_1,\phi_0,\phi_{-1})$.
Here,  we briefly  {summarize} the defining characteristics of each phase and discuss how we identify these phases in our HF calculations (for more details on the definition and properties of these phases, see Ref.~\cite{Ueda2010R}).
 
\vspace*{0.2cm}\noindent  {\bf  Ferromagnetic phase (F)}:   the condensate order parameter is of the form $(\sqrt{\tnc},0,0)$ for $p>0$. In this phase the condensate is fully magnetized along the direction of the applied field, i.e.,
\begin{equation}F_{\perp}^{\rm{c}}=0,\quad \mbox{and}\quad F_z^{\rm{c}}/n^{\rm{c}}=1,\label{Fcond}\end{equation}
where $F_{\perp}^{\rm{c}}=[(F_{x}^{\rm{c}})^2+(F_{y}^{\rm{c}})^2]^{1/2}$ is the transverse spin density.

\vspace*{0.2cm}\noindent   {\bf Antiferromagnetic phase (AF)}:   the condensate order parameter is of the form $(\sqrt{\tnc_{1,1}},0,\sqrt{\tnc_{-1,-1}})$.
 In this phase the condensate is partially magnetized along the direction of the applied field, i.e.,
\begin{equation}
F_{\perp}^{\rm{c}}=0,\quad \mbox{and}\quad0<F_z^{\rm{c}}/n^{\rm{c}}<1 .\label{AFcond}
\end{equation}

\vspace*{0.2cm}\noindent  {\bf Polar phase (P)}:  the condensate order parameter is of the form $(0,\sqrt{\tnc},0)$. In this phase the condensate is unmagnetized, i.e.,
\begin{equation}
F_{\perp}^{\rm{c}}=0,\quad \mbox{and}\quad F_z^{\rm{c}}/n^{\rm{c}}=0.\label{Pcond}
\end{equation}

\vspace*{0.2cm}\noindent  {\bf Broken-axisymmetry  phase (BA)}: the condensate order parameter is of the form $(\sqrt{\tnc_{1,1}},\sqrt{\tnc_{0,0}},\sqrt{\tnc_{-1,-1}})$ (see Appendix \ref{FHF} and Ref.~\cite{Ueda2010R} for more details).
In this phase the condensate is partially magnetized but tilts against the direction of the applied field, i.e.,
\begin{equation}
F_{\perp}^{\rm{c}}>0.\label{BAcond}
\end{equation}

We use the conditions (\ref{Fcond})-(\ref{BAcond}) to identify the phase of any self-consistent solution we obtain to the HF equations. Obtaining precise equality is not possible in finite precision numerical calculations and in practice we identify each phase when the appropriate equality (or inequality) is satisfied to one part in $10^{4}$ (e.g.~we identify the ferromagnetic phase by requiring   $F_z^{\rm{c}}/n^{\rm{c}}\ge0.9999$). 
 
\subsection{Antiferromagnetic interactions}

\subsubsection{Numerical results}
\begin{figure}[ht]
\includegraphics[width=0.9\linewidth]{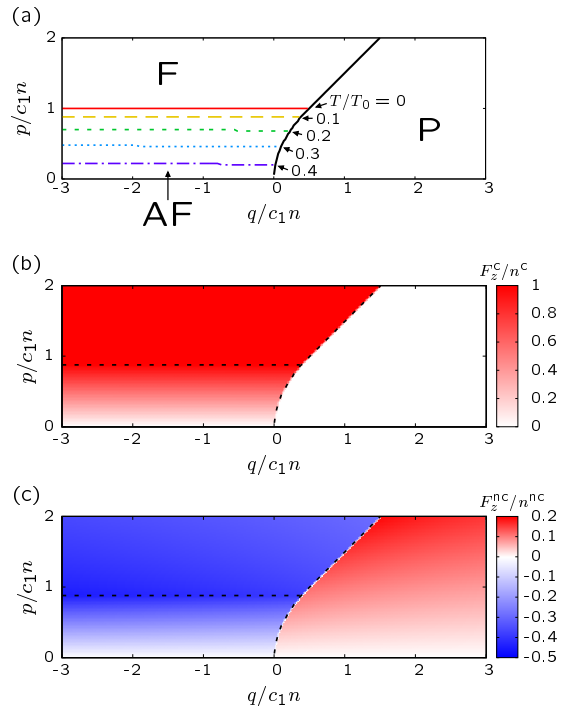}
\caption{(Color)
Results of the HF calculation for  antiferromagnetic interactions with $c_1/c_0=0.05$.
(a) Temperature dependence of the phase diagram in  $(q,p)$ space,
where the F--P and AF--P phase boundaries are independent of temperature.
The region of the AF phase shrinks as the temperature increases.
The longitudinal magnetization per atom of (b) the condensate and (c) the  non-condensate  at $T/T_0=0.1$.
The transverse magnetizations are always zero for both condensed and non-condensed atoms.
}
\label{fig:c1=0.05}
\end{figure}

The results for $c_1/c_0=0.05$ are summarized in Fig.~\ref{fig:c1=0.05}.
Figure~\ref{fig:c1=0.05}(a) shows the temperature dependence of the $q$--$p$ phase diagram.
The region of the P phase is unchanged, whereas the AF--F phase boundary moves downward as temperature increases.
Figures~\ref{fig:c1=0.05}(b) and \ref{fig:c1=0.05}(c) are the plots of the longitudinal magnetizations of condensate and non-condensate, respectively, at $T/T_0=0.1$.
When the condensate is in the P phase, the non-condensate is magnetized in the $z$ direction due to the linear Zeeman effect.
On the other hand, when the condensate is magnetized in the $z$ direction (i.e., in the F and AF phases), the non-condensate is magnetized anti-parallel to the condensate.
{This is because the condensate mainly occupies the lowest Zeeman sub-level ($i = 1$)  in these phases, and therefore, the residual non-condensate atoms prefer to populate the other spin states. This can be understood 
 as follows: The non-condensed atoms in spin states different from the condensate interact with the condensate only via the direct (Hartree) term; in contrast,  it is of higher energetic cost for non-condensate atoms to  occupy the same spin state as the condensate  because  both the direct (Hartree) and exchange (Fock) terms contribute.}

\subsubsection{AF--F phase boundary} 
Here, we focus on the temperature dependence of the linear Zeeman energy, $p_{\rm{b}}$, that specifies the AF--F phase boundary. 
The order parameter for the AF phase is given by
\begin{align}
 \begin{pmatrix}\phi_1 \\ \phi_0 \\ \phi_{-1}\end{pmatrix}
 =  \begin{pmatrix} \sqrt{n^{\rm c}_{1,1}} \\ 0 \\ \sqrt{n^{\rm c}_{-1,-1}} \end{pmatrix},
\label{eq:AF_nc}
\end{align}
where we can choose $\phi_{\pm 1}$ as positive real numbers without loss of generality,
because the phases of $\phi_{\pm 1}$ can be removed by
a gauge transformation and a spin rotation about the $z$ axis.
In other words, both the gauge transformation and spin rotation symmetries are spontaneously broken in the AF phase.
Since $\tnc_{ij}$ has the off-diagonal elements $n^{\rm c}_{1,-1}=n^{\rm c}_{-1,1}= \sqrt{n^{\rm c}_{1,1}n^{\rm c}_{-1,-1}}$,
$\tnnc_{ij}$, in general, has the off-diagonal components:
\begin{align}
{\bm n}^{\rm nc} = \begin{pmatrix}  n^{\rm nc}_{1,1} & 0 & (n^{\rm nc}_{-1,1})^* \\ 0 &  n^{\rm nc}_{0,0} & 0 \\
n^{\rm nc}_{-1,1} & 0 &  n^{\rm nc}_{-1,-1} \end{pmatrix}.
\label{eq:AF_nnc}
\end{align}

The generalized GPE~\eqref{genGPEu} for the AF phase
reduces to
\begin{align}
\begin{pmatrix} -\tp -\tmu & C_- n^{\rm nc}_{-1,1}\\
C_-(n^{\rm nc}_{-1,1})^*& \tp -\tmu \end{pmatrix}
\begin{pmatrix}\phi_1 \\ \phi_{-1} \end{pmatrix} = 0,
\label{eq:AF_cond}
\end{align}
where
\begin{align}
\tmu &= \mu - \left(q+c_0 n +c_1  n^{\rm nc}_{0,0} + C_+\frac{n^{\rm nc}_{1,1}+n^{\rm nc}_{-1,-1}}{2}\right),\\
\tp &= p - c_1 F^{\rm c}_z -\frac{c_0+3c_1}{2}F_z^{\rm nc}, \label{eq:def_tildep}\\
C_\pm &=c_0 \pm c_1, \label{eq:def_Cpm}
\end{align}
with $n=n^{\rm c}+n^{\rm nc}$.
At $T=0$, Eq.~\eqref{eq:AF_cond} has an AF solution ($\phi_{\pm1}\neq 0$)
when $\tilde{p}=0$, that is, $p=c_1 F_z^{\rm c}$.
From the fact that $F_z^{\rm c}=n^{\rm c}$ at the AF--F phase boundary,
$p_{\rm b}$ at $T=0$ is given by
\begin{align}
\frac{p_{\rm b}}{c_1n}=1.
\label{pbT0}
\end{align}

At $T\neq 0$, the condition that Eq.~\eqref{eq:AF_cond} has a nontrivial solution determines $\tmu$.
Substituting $\tmu$ and the solution of $(\phi_1, \phi_{-1})$ to the HF equations and solving self-consistently,
we obtain the following relation among $p$, $F_z^{\rm c}$, and $F_z^{\rm nc}$ in the AF phase:
\begin{align}
p =& \frac{3C_+-2C_-}{4}F^{\rm c}_z + \frac{4C_+-3C_-}{4}F^{\rm nc}_z \nonumber\\
&- \frac{1}{2}\sqrt{\left(\frac{C_+ F^{\rm c}_z-C_-F^{\rm nc}_z}{2}\right)^2 + C_-^2 F^{\rm c}_z F^{\rm nc}_z}.
\label{eq:p-AF}
\end{align}
The detailed derivation of Eq.~\eqref{eq:p-AF} is given in Appendix~\ref{sec:AF-F}.
Using the fact that $F_z^{\rm c}=n^{\rm c}$ at the AF--F phase boundary,
 {and expanding Eq.~\eqref{eq:p-AF} in terms of $F_z^{\rm nc}/n^{\rm c}$,}
the phase boundary is approximated as
\begin{align}
 \frac{p_{\rm b}}{c_1 n} \cong \frac{n^{\rm c}}{n} + \frac{3c_0+c_1}{c_0+c_1}\frac{F^{\rm nc}_z}{n}.
\label{eq:AF-F_boundary}
\end{align}
The right-hand side of Eq.~\eqref{eq:AF-F_boundary} goes to unity as $T\to 0$,
being consistent with Eq.~\eqref{pbT0}.
The first term on the right-hand side of Eq.~\eqref{eq:AF-F_boundary} describes the shift in the boundary due to the thermal depletion of the condensate,
while the second term describes the interaction of the non-condensed component back on the condensate and acts to reduce the value of $p_{\rm b}$ since $F_z^{\rm nc}<0$ [see Fig.~\ref{fig:c1=0.05}(c)].

This result can be understood in terms of two underlying effects that compete against each other:\\
(i) The non-condensate magnetization $F^{\rm nc}_z$ increases the effective linear Zeeman energy [see Eq.~\eqref{eq:def_tildep}], i.e.,~increases the energy difference between the $i=1$ and $-1$  components of the condensate [see Eq.~\eqref{eq:AF_cond}].  This causes  $|\phi_1|$   to increase relative to $|\phi_{-1}|$, and thus tends to reduce the value of $p_{\rm b}$ at the phase boundary  (where $\phi_{-1}=0$).\\
(ii) The non-condensate spin coherence plays a nontrivial role through  exchange (Fock) collisions between condensate and non-condensate atoms of the type $(i,{\bf 0})+(j,{\bf k}) \leftrightarrow (i,{\bf k})+(j,{\bf 0})$:  Off-diagonal elements of $n^{\rm nc}_{ij}$ contribute to enhancing the coupling between $\phi_1$ and $\phi_{-1}$ [see Eq.~\eqref{eq:AF_cond}], thus acting to make 
$|\phi_1|$ and $|\phi_{-1}|$  more similar, and hence supporting the AF phase (i.e.,~this effect tends to increase $p_{\rm b}$).

To quantify the competition between these two effects we neglect the non-condensate spin coherence  by explicitly setting $\tnnc_{-1,1}=0$ in Eq.~(\ref{eq:AF_nnc}) {and calculate $p_{\rm b}$}. In such a case, Eq.~\eqref{eq:AF_cond} has an AF solution when {$\tp=0$},
resulting in
\begin{align}
 \frac{p_{\rm b}}{c_1n}= \frac{n^{\rm c}}{n} + \frac{c_0+3c_1}{2c_1}\frac{F^{\rm nc}_z}{n}.
\label{eq:AF-F_boundary_diagonal}
\end{align}
The larger pre-factor of the last term demonstrates  that  when non-condensate spin coherence is neglected [i.e.,~only effect (i) contributes] the phase bound  $p_{\rm b}$ is more significantly reduced.

\begin{figure}[ht]
\includegraphics[width=0.9\linewidth]{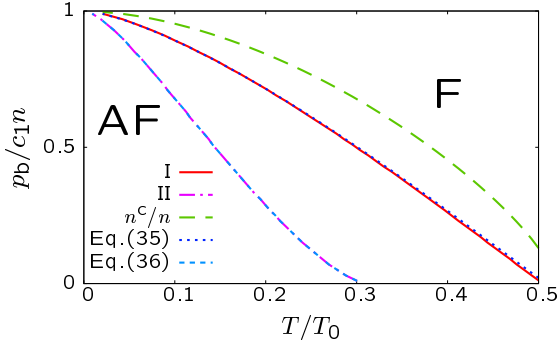}
\caption{(Color online)
Temperature dependence of the AF--F phase boundary $p_{\rm b}$ at $q=-3c_1n$ and $c_1/c_0=0.05$
obtained by (I) the full-HF calculation and (II) the HF calculation but neglecting the off-diagonal elements of $\tnnc_{ij}$,
together with the curves indicating $n^{\rm c}/n$, Eq.~\eqref{eq:AF-F_boundary}, and Eq.~\eqref{eq:AF-F_boundary_diagonal}.
}
\label{fig:pb}
\end{figure}

Figure~\ref{fig:pb} shows the temperature dependence of $p_{\rm b}$  for a particular choice of the quadratic Zeeman energy ($q=-3c_1n$) obtained by the full HF calculation (I)
and the HF calculation with  the off-diagonal elements of $\tnnc_{ij}$ neglected (II)~\footnote{Irrespective of whether off-diagonal parts of $\tnnc_{ij}$ arise, we only include the diagonal parts when evaluating the self-consistent HF Hamiltonian [see Eq.~(\ref{Aij})].},
which show good agreement with Eqs.~\eqref{eq:AF-F_boundary} and \eqref{eq:AF-F_boundary_diagonal}, respectively.
The deviations of the curves I and II from $p_{\rm b}/(c_1n)=n^{\rm c}/n$ are the effects of the thermal components ($F^{\rm nc}_z/n$).
Note that the value of  $p_{\rm b}$  significantly decreases when we neglect the spin coherence of the non-condensate.
We find that the temperature dependence of the condensate fraction $n^{\rm c}/n$ and the non-condensate magnetization $F^{\rm nc}_z$ are almost the same for I and II,
so the difference in the phase boundaries arises from the coefficients of $F^{\rm nc}_z/n$ in Eqs.~\eqref{eq:AF-F_boundary} and \eqref{eq:AF-F_boundary_diagonal}.
For the case of the full HF calculation [Eq.~\eqref{eq:AF-F_boundary}], $p_{\rm b}$ is insensitive to the value of $c_1$ as long as $c_1/c_0\ll 1$.
On the other hand, Eq.~\eqref{eq:AF-F_boundary} is strongly dependent on $c_1/c_0$, in particular when $c_1/c_0$ is small.
We have also numerically calculated $p_{\rm b}$ for the interaction parameters of $c_1/c_0=0.005$ and $0.5$.
The results agree with Eqs.~\eqref{eq:AF-F_boundary} and \eqref{eq:AF-F_boundary_diagonal}.

\subsection{Ferromagnetic interactions}

\subsubsection{Numerical Results}
\begin{figure}[ht]
\includegraphics[width=0.9\linewidth]{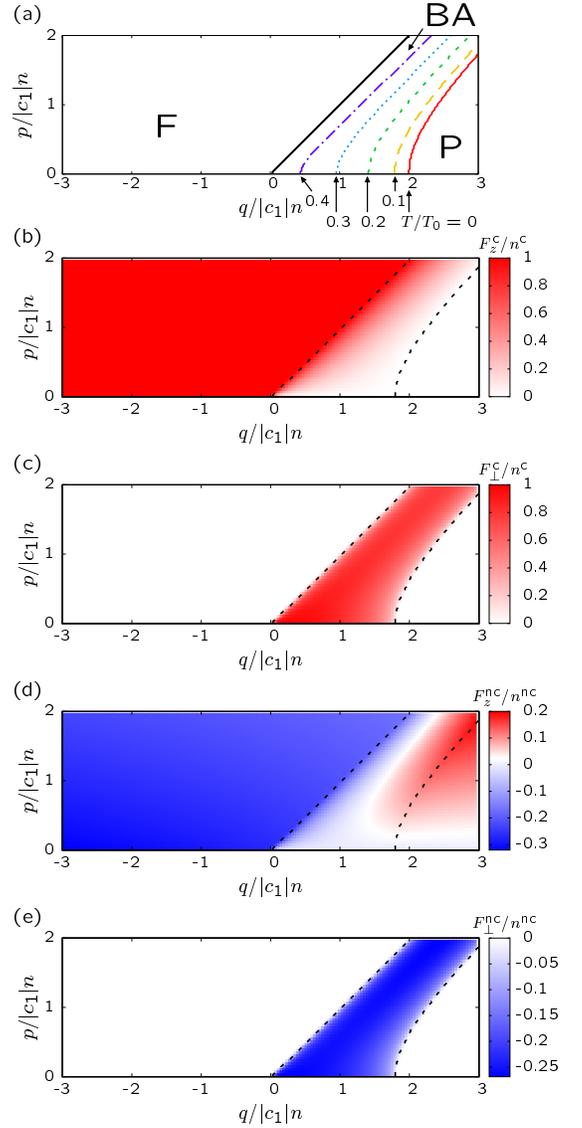}
\caption{(Color)
Results of the HF calculation for   ferromagnetic interactions  with $c_1/c_0=-0.05$.
(a) Temperature dependence of the phase diagram in $(q,p)$ space,
where the F--BA phase boundary is independent of temperature.
The region of the BA phase shrinks as the temperature increases.
The longitudinal and transverse magnetization per atom   at $T/T_0=0.1$ for (b), (c) the condensate 
and (d), (e) the non-condensate.
In (e), $F^{\rm nc}_{\perp}<0$ means that 
the transverse magnetization of the non-condensate is anti-parallel to that of the condensate.
}
\label{fig:c1=-0.05}
\end{figure}

The numerical results for $c_1/c_0=-0.05$ are summarized in Fig.~\ref{fig:c1=-0.05}.
Figure~\ref{fig:c1=-0.05}(a) shows the temperature dependence of the $q$--$p$ phase diagram.
The region of the F phase is unchanged, whereas the BA--P phase boundary moves to the left-hand side as temperature increases. 
Figures~\ref{fig:c1=-0.05}(b) and \ref{fig:c1=-0.05}(c) are the plots of the longitudinal and transverse magnetizations of condensate atoms, respectively, at $T/T_0=0.1$,
and Figs.~\ref{fig:c1=-0.05}(d) and \ref{fig:c1=-0.05}(e) show the same quantities for the non-condensate.
In Fig.~\ref{fig:c1=-0.05}(e), $F_{\perp}^{\rm nc}<0$ means that the transverse magnetization of the non-condensate is anti-parallel to that of the condensate. 
{As in the case of the AF and F phases of Fig.~\ref{fig:c1=0.05}, the non-condensate magnetization is roughly anti-parallel to that of the condensate, except for the vicinity of the BA--P phase boundary where the condensate magnetization becomes small.}

\subsubsection{BA--P phase boundary} 
We investigate the temperature dependence of the BA--P phase boundary $q_{\rm b}$ at $p=0$.
In the BA phase at $p=0$ the condensate magnetization is purely transverse
and vanishes at $q=q_{\rm b}$.
Note that the numerical result [Fig.~\ref{fig:c1=-0.05}(e)] shows that the non-condensed component is also magnetized in the transverse direction (see also Ref.~\cite{Phuc2011a}), indicating the existence of the spin coherence in the non-condensate.
This is because the spin rotational symmetry about the $z$ axis is broken in the HF Hamiltonian~\eqref{KHF} due to the existence of the transversely magnetized condensate.

At $T=0$, the BA--P phase boundary is given by~\cite{Ueda2010R}
\begin{align}
 \frac{q_{\rm b}}{|c_1|n} = 2.
\end{align}
At finite temperature, 
by solving the Gross-Pitaevskii and HF equations self-consistently,
we obtain the following relation for BA--P boundary (see Appendix~\ref{FHF} for the derivation):
\begin{align}
 \frac{q_{\rm b}}{|c_1|n} &\cong 2\frac{n^{\rm c}}{n} - \frac{4(3c_0-5|c_1|)}{c_0-|c_1|}\frac{\dnc}{n},
\label{eq:BA-P_boundary}
\end{align}
where
\begin{align}
\dnc=\frac{1}{2}\left(n^{\rm nc}_{1,1}-n^{\rm nc}_{0,0}+n^{\rm nc}_{-1,1}\right).
\label{eq:def_dnc_BA}
\end{align}

As in the case of antiferromagnetic interactions,
the non-condensate spin coherence has a significant effect on the location of the phase boundary.
If we neglect the off-diagonal elements of $\tnnc_{ij}$, the phase boundary is changed to
\begin{align}
\frac{q_{\rm b}}{|c_1|n}= 2\frac{n^{\rm c}}{n} - \frac{c_0+|c_1|}{|c_1|}\frac{\dnc}{n},
\label{eq:BA-P_boundary_diagonal}
\end{align}
where $\dnc$ is defined in Eq.~\eqref{eq:def_dnc_BA} but with $n^{\rm nc}_{1,-1}=0$.
The derivation of Eq.~\eqref{eq:BA-P_boundary_diagonal} is given in Appendix~\ref{FHF2}.

Figure~\ref{fig:qb} shows the temperature dependence of $q_{\rm b}$ at $p=0$
obtained  by the full HF calculation (I)
and the HF calculation with the off-diagonal elements of $\tnnc_{ij}$ neglected (II),
which show good agreement with Eqs.~\eqref{eq:BA-P_boundary} and \eqref{eq:BA-P_boundary_diagonal}, respectively.
The deviations of the curves I and II from $q_{\rm b}/(|c_1|n)=2n^{\rm c}/n$ are the effects of the non-condensate ($\dnc/n$).
As in the case of Fig.~\ref{fig:pb}, $q_{\rm b}$ is greatly suppressed when we neglect the coherence of the non-condensate.
The difference also comes from the coefficients of $\dnc/n$ in Eqs.~\eqref{eq:BA-P_boundary} and \eqref{eq:BA-P_boundary_diagonal}:
Eq.~\eqref{eq:BA-P_boundary} is insensitive to the value of $c_1$ as long as $|c_1|/c_0\ll 1$;
while Eq.~\eqref{eq:BA-P_boundary} is strongly dependent on $c_1/c_0$, in particular when $|c_1|/c_0$ is small.
We have also numerically calculated $q_{\rm b}$ for the interaction parameters of $c_1/c_0=-0.005$ and $-0.5$.
The results agree with Eqs.~\eqref{eq:BA-P_boundary} and \eqref{eq:BA-P_boundary_diagonal}.

\begin{figure}[ht]
\includegraphics[width=0.9\linewidth]{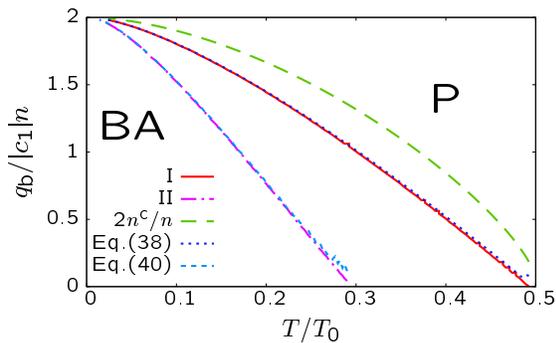}
\caption{
Temperature dependence of the BA--P phase boundary $q_{\rm b}$ at $p=0$ and $c_1/c_0=-0.05$
obtained by (I) the full-HF calculation and (II) the HF calculation but neglecting the off-diagonal elements of $\tnnc_{ij}$,
together with the curves indicating $2n^{\rm c}/n$, Eq.~\eqref{eq:BA-P_boundary}, and Eq.~\eqref{eq:BA-P_boundary_diagonal}.
}
\label{fig:qb}

\end{figure}

The interpretation of the above results is the similar to the case of antiferromagnetic interactions.
In Eq.~\eqref{eq:def_dnc_BA} the main contribution to $d^{\rm nc}$ comes from the population difference between $i=1$ and $0$ components, $n^{\rm nc}_{1,1}-n^{\rm nc}_{0,0}$ ($=n^{\rm nc}_{-1,-1}-n^{\rm nc}_{0,0}$ for $p=0$), which
induces an energy difference between condensed atoms in the $i=0$ and $\pm1$ components via the exchange (Fock) terms [the last terms in the first and second lines of Eq.~\eqref{Liju}]. 
Hence, $d^{\rm nc}$ contributes to increasing $|\phi_0|$ relative to $|\phi_{\pm1}|$, and thus tends to reduce the value of $q_{\rm b}$.
On the other hand, the off-diagonal elements of $n^{\rm nc}_{ij}$, in particular $n^{\rm nc}_{\pm1,0}$ and $n^{\rm nc}_{0,\pm1}$,  compete against this by coupling condensate atoms in $i=0$ and $\pm1$ states [see Eq.~\eqref{Liju}], which acts to 
balance the condensate population in these states and strengthen the BA phase (i.e.,~this effect tends to increase $q_{\rm b}$).

\section{Conclusions and outlook}
In this work  we  have formulated a self-consistent HF theory to characterize the phase diagram of a spin-1 Bose gas at finite temperature. Numerical results, presented over a wide parameter regime, show that certain phase boundaries change appreciably with temperature. We have developed analytic results that accurately describe these shifts in phase boundaries as a function of the interaction parameters and the properties of the non-condensate. 

  Our treatment includes spin coherence for the non-condensate component of the system,  
which naturally develops via coherent collisions with the condensate.
 Our calculations show that  
{the non-condensate spin coherence is crucial to stabilizing the AF and BA phases, in which the spin rotational symmetry is spontaneously broken.
Indeed, neglecting spin coherence in the thermal cloud leads to  significant shifts in the locations of the phase boundaries from the full HF calculations.
  
 The effect of the thermal fluctuations on the condensate order is a key prediction that could be explored in experiments. Early measurements made by the MIT group \cite{Stenger1999a} mapped out parts of the phase diagram using a  $^{23}$Na condensate (with antiferromagnetic interactions). In that work the temperature of the system was estimated to be about 100 nK, sufficiently hot that thermal effects should be relevant, however they measured the P--AF phase boundary which we predict to be temperature insensitive [see Fig.~\ref{fig:c1=0.05}(a)]. Aided by improvements in techniques for measuring spinor gas properties (e.g.~see Refs.~\cite{Higbie2005a,Black2007a}) it should be feasible to precisely determine the finite temperature phase diagram in experiments and compare to our predictions. 

It would also be interesting to experimentally investigate the role of the non-condensate spin coherence. Our results show that a large change in the phase boundary occurs when the non-condensate  coherence is removed  (see Figs.~\ref{fig:pb} and \ref{fig:qb}). Given the large difference in the decoherence times for the  {condensate and non-condensate} spin coherence \cite{Higbie2005a} it may be possible to use external fields to reduce (or remove) the spin coherence of the non-condensate, yet leave that of the condensate intact. In the vicinity of the phase boundary this could allow the condensate to exist in a metastable state which would transition to a new phase as the  non-condensate spin coherence is eventually reestablished.

 On the theoretical front many challenges and opportunities exist for extending our understanding of spinor gases beyond the HF approximation. A natural extension is to develop a quasi-particle based mean-field theory such as the HF-Bogoliubov-Popov formalism \cite{Isoshima2000a,Phuc2011a}. In Ref.~\cite{Phuc2011a} we applied this theory  to compute the BA--P phase boundary as a function of temperature for $p=0$ and the parameters of $^{87}$Rb. The predictions of  {Ref.~\cite{Phuc2011a}} are quantitatively similar to the HF results we present here, with the notable exception of the $T\to0$ limit where we have found that the quantum depletion (excluded in the HF theory) acts to increase $q_{\rm b}$ to a value greater that $2|c_1|\tnc$.
An alternative direction is the use of classical field techniques \cite{cfieldRev2008} which, within their regime of validity, will provide a dynamical description of the finite temperature spinor system, and have already seen some initial applications to quasi-two-dimensional spinor gases \cite{Pietila2010a,*Su2011a}.
 Another avenue for consideration is the inclusion of dipole-dipole interactions between atoms   into the finite temperature description (e.g.~see \cite{Ronen2007a,*Bisset2011a}).  These long-range interactions have been predicted to give rise to interesting new features in the ground state phase diagram \cite{Kawaguchi2006a}, and are thought to be important for explaining some of the observations in the $^{87}$Rb spinor gas \cite{Vengalattore2008a}.  

 \section*{Acknowledgements}
 YK and NTP were supported by KAKENHI (22340114, 22740265, 22103005), a Global COE Program \textquotedblleft the Physical Sciences Frontier\textquotedblright, and the Photon Frontier Network Program, from MEXT of Japan, and by JSPS and FRST under the Japan-New Zealand Research Cooperative Program. 
PBB was supported by Marsden contract UOO0924 and FRST IIOF contract UOOX0915.

\appendix
\section{Hartree-Fock theory and thermodynamic parameters}\label{AppdHF}
The HF theory can be derived  {by assuming that} the many-body density matrix is given by
\begin{equation}
D_0=\frac{1}{Z_0}e^{-\beta \hat K_{\rm{HF}}},
\end{equation}
where  $Z_0=\rm{Tr}\{e^{-\beta \hat K_{\rm{HF}}}\}$ and  $\hat K_{\rm{HF}}=\int d\r \sum_{i,j}\hat{\delta}^\dagger_iA_{ij}\hat{\delta}_j$ is  the assumed single particle form for the HF Hamiltonian.  The variational principle applied to determine $\hat K_{\rm{HF}}$  {(or, equivalently, $A_{ij}$)} is that $D_0$ makes the thermodynamic potential $\Phi(D)$ stationary, where
\begin{equation}
\Phi(D)=\mathrm{Tr}\{k_{\rm B}TD\ln D+D\hat{\mathcal{H}}-\mu D\hat{N}\},\label{ThermPot}
\end{equation}
with $\hat{N}$  {being} the number operator. This procedure gives  the form of the Gross-Pitaevskii and HF equations used  {in this paper} [{i.e.,} Eqs.~(\ref{genGPE}), (\ref{KHF}) and (\ref{AHF})].

In terms of the self-consistent solution of the HF equations thermodynamic parameters can be evaluated. The HF energy is given by
\begin{align}
E_{\rm{HF}}=& 
 {\mathrm{Tr}\{D_0\hat{\mathcal{H}}\}}\\
=&\int d\mathbf{r}\left\{\sum_{j}\,\left[\phi_{j}^{*}(h_{0})_{jj}\phi_{j}+\sum_{\lambda}\bar{n}_{\lambda}u_{j}^{\lambda*}(h_{0})_{jj}u_{j}^{\lambda}\right]\right.\nonumber\\
&+\frac{c_{0}}{2}\left[(n^{{\rm c}}+n^{{\rm nc}})^2 +\sum_{ij}\tnnc_{ij}(2\tnc_{ {ji}}+\tnnc_{ {ji}})\right]\nonumber\\ 
&+\sum_{\alpha}
\frac{c_{1}}{2} \Bigg[(F_{\alpha}^{{\rm c}}+F_{\alpha}^{n{\rm c}})^2  \nonumber \\
&\left.  + \sum_{ijkl}(f_{\alpha})_{ij}(f_{\alpha})_{kl}\,\tnnc_{kj}\left(2\tnc_{il}  +\tnnc_{il}\right)\Bigg] \right\},
\label{EHF}
\end{align}
and by evaluating Eq.~(\ref{ThermPot}), using the self-consistently determined HF density matrix, the free energy of the HF solution ($\Phi_{\rm{HF}}$) can be determined. Equivalently it can be evaluated as
\begin{equation}
\Phi_{\rm{HF}}=E_{\rm{HF}}-\mu N-TS_{\rm{HF}},
\end{equation}
where the entropy is 
\begin{align}
S_{\rm{HF}}  &=-k_{B}\sum_{\lambda}[\bar{n}_{\lambda}\ln\bar{n}_{\lambda}-(1+\bar{n}_{\lambda})\ln(1+\bar{n}_{\lambda})].
\end{align}

\section{Derivation of Eq.~\eqref{eq:AF-F_boundary}}
\label{sec:AF-F}
 
In this and the following appendixes, we use bold quantities to represent matrix quantities for notational efficiency, for example $\tnc_{ij}\to\nc$, $\tnnc_{ij}\to\nnc$, and $\delta_{ij}\to \bm 1$.
We also introduce $\dc\equiv n^{\rm c}_{1,-1}=n^{\rm c}_{-1,1}$, $\dnc \equiv n^{\rm nc}_{-1,1}$, $n^{\rm c}_i \equiv n^{\rm c}_{ii}$, and $n^{\rm nc}_i \equiv n^{\rm nc}_{ii}$.

We start from Eq.~\eqref{eq:AF_cond}.
From the condition that Eq.~\eqref{eq:AF_cond} has a nontrivial solution, $\tmu$ is obtained as
\begin{align}
\tmu =  \pm \sqrt{\tp^2+C_-^2|\dnc|^2}.
\label{eq:tmu}
\end{align}
Choosing the lower chemical potential, the order parameter is given by
\begin{subequations}
\begin{align}
\phi_1    &= \sqrt{\frac{n^{\rm c}}{2} \left(1+\frac{\tp}{\lambda}\right)},\\
\phi_{-1} &= -e^{-i\theta}\sqrt{\frac{n^{\rm c}}{2} \left(1-\frac{\tp}{\lambda}\right)},
\end{align}
\label{eq:keron}
\end{subequations}
where
\begin{align}
\lambda &= \sqrt{\tp^2+C_-^2|\dnc|^2},\\
\theta &= {\rm arg}(\dnc).
\end{align}
Since we have chosen $\phi_{\pm 1}$ to be positive real numbers,
$\dnc$ is a negative number ($\theta=\pi$).
From Eq.~\eqref{eq:keron}, we obtain the relation between the condensate spin density and $\dnc$:
\begin{align}
 \frac{\dc}{F_z^{\rm c}} = \frac{\phi_1\phi_{-1}}{|\phi_1|^2-|\phi_{-1}|^2}
= - \frac{C_-\dnc}{2\tp}.
\label{eq:dc}
\end{align}

Next, by substituting Eqs.~\eqref{eq:AF_nc} and \eqref{eq:AF_nnc} to Eq.~(\ref{Aij}),
we obtain
\begin{align}
 \mathcal{\bm A}&= \left(-\mu + c_0 n\right)\bm 1 +
\begin{pmatrix} -p+c_1 F_z + q & 0 & 0 \\ 0 & 0 & 0 \\ 0 & 0 & p-c_1 F_z +q \end{pmatrix}\nonumber\\
&+\begin{pmatrix} 
C_+ n_1 + c_1n_0 & 0 & C_- d \\ 0 & c_0 n_0 + c_1(n_1+n_{-1}) & 0 \\
C_- d & 0 & C_+ n_{-1} + c_1n_0 \end{pmatrix},
\end{align}
where $n_i=n^{\rm c}_i+ n^{\rm nc}_i$, $F_z=F_z^{\rm c}+F_z^{\rm nc}$, and $d=\dc+\dnc$.
The eigenvalue for the $i=0$ component is immediately obtained as
\begin{align}
\kappa_{0} = c_0 n +c_0 n_0 + c_1(n_1+n_{-1}) -\mu.
\end{align}
For the $i=\pm1$ components, we need to diagonalize the following $2\times 2$ matrix:
\begin{align}
\tilde{\mathcal{\bm A}}
=&
\left[-\tmu + c_1 n^{\rm c}_0+ \frac{C_+}{2}(n^{\rm c}_1+n^{\rm c}_{-1})\right]\bm 1\nonumber\\
& +
\begin{pmatrix}
 -\tp + C_+ F_z^{\rm c}/2 & C_-d \\
 C_-d & \tp - C_+ F_z^{\rm c}/2
\end{pmatrix}.
\label{eq:tildeH_HF}
\end{align}
This matrix is almost the same as Eq.~\eqref{eq:AF_cond}, and the eigenvalues and eigenvectors are given by
\begin{align}
\kappa_\pm =&  - \tmu + c_1 n^{\rm c}_0 +  \frac{C_+}{2}\left(n_1^{\rm c}+n_{-1}^{\rm c}\right)
\pm \lambda'\\
\begin{pmatrix} \bar{u}_{1}^{(\pm)} \\ \bar{u}_{-1}^{(\pm)} \end{pmatrix}
 =&\frac{1}{\sqrt{2\lambda'}}
\begin{pmatrix} 
\sqrt{\lambda'\mp (\tp-C_+F_z^{\rm c}/2)} \\ 
\pm {\rm sgn}(d) \sqrt{\lambda'\pm (\tp-C_+F_z^{\rm c}/2)}
\end{pmatrix},
\end{align}
where
\begin{align}
\lambda' \equiv& \sqrt{(\tp - C_+ F_z^{\rm c}/2)^2+C_-^2 d^2}.
\end{align}
Since $n^{\rm nc}_{\pm 1}$ and $d^{\rm nc}$ are self-consistently determined so as to satisfy Eq.~\eqref{eq:tnnc_Li},
we obtain the relation between $F_z^{\rm nc}$ and $\dnc$:
\begin{align}
 \frac{\dnc}{F_z^{\rm nc}} 
&= \frac{\displaystyle\sum_{\nu=\pm}\bar{u}_{-1}^{(\nu)*}\bar{u}_1^{(\nu)}
\textrm{Li}_{3/2}(e^{-\beta\kappa_\nu})}
{\displaystyle\sum_{\nu=\pm}\left[\bar{u}_1^{(\nu)*}\bar{u}_1^{(\nu)}-\bar{u}_{-1}^{(\nu)*}\bar{u}_{-1}^{(\nu)}\right]
\textrm{Li}_{3/2}(e^{-\beta\kappa_\nu})}
 \nonumber\\
&= -\frac{C_-d}{2\tp - C_+F_z^{\rm c}}.
\label{eq:dnc}
\end{align}

Equations~\eqref{eq:dc} and \eqref{eq:dnc} are rewritten as a linear equation of $\dc$ and $\dnc$:
\begin{align}
 \begin{pmatrix} 
  2\tp  & C_- F^{\rm c}_z \\ C_- F^{\rm nc}_z & 2\tp - C_+F^{\rm c}_z + C_-F^{\rm nc}_z
 \end{pmatrix}
 \begin{pmatrix} \dc \\ \dnc \end{pmatrix} =0.
\end{align}
In order for $\dc$ and $\dnc$ to have a non-trivial solution, $\tp, F^{\rm c}_z$ and $F^{\rm nc}_z$ have to satisfy
\begin{align}
2\tp (2\tp - C_+F^{\rm c}_z + C_-F^{\rm nc}_z)  -C_-^2 F^{\rm c}_zF^{\rm nc}_z = 0,
\label{eq:p}
\end{align}
Solving Eq.~\eqref{eq:p} in terms of $p$, we obtain Eq.~\eqref{eq:p-AF},
where we have chosen the sign in front of the square root term so that Eq.~\eqref{eq:p-AF} continuously goes to the solution at $T=0$.

\section{Derivation of Eqs.~\eqref{eq:BA-P_boundary} and \eqref{eq:BA-P_boundary_diagonal}}
\label{sec:BA-P}
In the BA phase at $p=0$ the condensate is magnetized in the transverse direction.
Without loss of generality, we can choose the direction of the magnetization in the $x$ direction, i.e.,
the magnetic field is applied in the $z$ direction and spontaneous magnetization arises in the $x$ direction.
We then move to the frame of reference which is rotated around the $y$ axis by $\pi/2$.
In this frame of reference, the magnetic field is applied in the $-x$ direction and the magnetization arises in the $z$ direction.
In this Appendix all results are given in this frame of reference unless specified otherwise.
The magnetic sub-level $i$ in the rotated frame corresponds to the eigenvalue of $f_x$ in the laboratory frame.

\subsection{Full-HF calculation}\label{FHF}
The matrices $\mathcal{\bm L}$ and $\mathcal{\bm A}$ in the rotated frame are given by
\begin{align}
 \mathcal{\bm L} =& q f_x^2+ c_0\left[n\bm 1 + ({\bm n}^{\rm nc})^{\rm T}\right] \nonumber\\
&+ c_1\sum_\alpha \left[ F_\alpha f_\alpha + f_\alpha ({\bm n}^{\rm nc})^{\rm T} f_\alpha\right], \label{eq:GP_BA}\\
\mathcal{\bm A} =& \mathcal{\bm L}  -\mu \bm 1 + c_0({\bm n}^{\rm nc})^{\rm T} + c_1 \sum_{\alpha} f_\alpha({\bm n}^{\rm c})^{\rm T}f_\alpha.\label{eq:L_BA}
\end{align}
Since the matrix elements of $f_x^2$ are given by
\begin{align}
f_x^2= \begin{pmatrix} 1/2 & 0 & 1/2 \\ 0 & 1 & 0 \\ 1/2 & 0 & 1/2 \end{pmatrix},
\end{align}
we can assume that $i=0$ and $i=\pm1$ components are decoupled:
\begin{align}
\nc &= \begin{pmatrix} n^{\rm c}_1 & 0 & \dc \\ 0 & 0 & 0 \\
\dc & 0 & n^{\rm c}_{-1} \end{pmatrix},\ \ 
\nnc = \begin{pmatrix}  n^{\rm nc}_1 & 0 & (\dnc)^* \\ 0 &  n^{\rm nc}_0 & 0 \\
\dnc & 0 &  n^{\rm nc}_{-1} \end{pmatrix}.
\label{eq:nncBA}
\end{align}
The order parameter for the BA phase in the laboratory frame is given by $\sqrt{n^{\rm c}/2}(a,\sqrt{2}b,a)^{\rm T}, (a,b\in \mathbb{R}, a^2+b^2=1, 0\le a \le 1/\sqrt{2})$ \cite{Ueda2010R},
which is transformed in the rotated frame as $\sqrt{n^{\rm c}/2}(a+b,0,a-b)^{\rm T}$.
It follows that $\dc$ is always negative in the BA phase because $0\le a \le 1/\sqrt{2}\le b \le 1$.
When the system is in the polar phase ($a=0, b=1$), we have $\dc = -n^{\rm c}/2$.

From Eq.~\eqref{eq:GP_BA}, the $i=\pm1$ components should satisfy
\begin{align}
\begin{pmatrix} \tp  -\tmu & q/2+C_-\dnc\\
q/2+C_-(\dnc)^*& -\tp -\tmu \end{pmatrix}
\begin{pmatrix}\phi_1 \\ \phi_{-1} \end{pmatrix} = 0,
\label{eq:BA_cond}
\end{align}
where
\begin{align}
\tmu &= \mu - \left(q/2+c_0 n +c_1  n^{\rm nc}_0 + C_+\frac{n^{\rm nc}_1+n^{\rm nc}_{-1}}{2}\right),\\
\tp &= c_1F_z^{\rm c}+\frac{c_0+3c_1}{2}F_z^{\rm nc},
\end{align}
and $C_{\pm}$ are defined in Eqs.~\eqref{eq:def_Cpm}.
From the condition that Eq.~\eqref{eq:BA_cond} has a nontrivial solution, $\tmu$ is determined as
\begin{align}
\tmu =
&\pm \sqrt{\tp^2+\left|\frac{q}{2}+C_-\dnc\right|^2}.
\end{align}
Choosing the lower chemical potential, 
the order parameter is given by
\begin{subequations}
\begin{align}
\phi_1    &= \sqrt{\frac{n^{\rm c}}{2} \left(1-\frac{\tp}{\lambda}\right)},\\
\phi_{-1} &= -e^{-i\theta}\sqrt{\frac{n^{\rm c}}{2} \left(1+\frac{\tp}{\lambda}\right)},
\end{align}
\label{eq:keron2}
\end{subequations}
where
\begin{align}
\lambda &\equiv \sqrt{\tp^2+\left(\frac{q}{2}+C_-\dnc\right)^2},\\
\theta &\equiv {\rm arg}\left(\frac{q}{2}+C_-\dnc\right).
\end{align}
Since $\phi_{-1}$ is assumed to be a negative real number, $\theta$ has to be zero, that is, $\dnc$ is real and satisfies $q/2+C_-\dnc>0$.
From Eq.~\eqref{eq:keron2}, we obtain the relation between $F_z^{\rm c}$ and $\dc$:
\begin{align}
 \frac{\dc}{F_z^{\rm c}} = \frac{q/2+C_-\dnc}{(C_+-C_-)F^{\rm c}_z+(2C_+-C_-)F^{\rm nc}_z}.
\label{eq:dc_BA}
\end{align}

Next, we consider the equation for the non-condensate part.
The matrix $\mathcal{\bm A}$ is given by
\begin{align}
 \mathcal{\bm A}=& \left(-\mu + c_0 n\right)\bm 1 +
\begin{pmatrix} c_1 F_z + q/2 & 0 & q/2 \\ 0 & q & 0 \\ q/2 & 0 & -c_1 F_z +q/2 \end{pmatrix}\nonumber\\
&+\begin{pmatrix} 
C_+n_1 + c_1n_0 & 0 & C_-d \\ 0 & c_0 n_0 + c_1(n_1+n_{-1}) & 0 \\
C_-d & 0 & C_+n_{-1} + c_1n_0 \end{pmatrix}.
\end{align}
For the $i=\pm 1$ components, we need to diagonalize the $2\times 2$ matrix:
\begin{align}
\tilde{\mathcal{\bm A}}
=&
\left(-\tmu + c_1 n^{\rm c}_0 + C_+\frac{n^{\rm c}_1+n^{\rm c}_{-1}}{2}\right)\bm 1 \nonumber\\
&+\begin{pmatrix}
 \tp + C_+ F^{\rm c}_z/2  & q/2+C_-d \\
 q/2+C_-d & -\tp - C_+F^{\rm c}_z/2
\end{pmatrix}.
\end{align}
The eigenvalues and eigenvectors of $\tilde{\mathcal{\bm A}}$ are given by
\begin{align}
\kappa_\pm =& - \tmu + c_1 n^{\rm c}_0 +  C_+\frac{n^{\rm c}_1+n^{\rm c}_{-1}}{2} 
\pm \lambda',\\
\begin{pmatrix} \bar{u}_{1}^{(\pm)} \\ \bar{u}_{-1}^{(\pm)} \end{pmatrix}
 =&\frac{1}{\sqrt{2\lambda'}}
\begin{pmatrix} 
\sqrt{\lambda'\pm (\tp+C_+F^{\rm c}_z/2)} \\ 
\pm e^{-i\theta'} \sqrt{\lambda'\mp (\tp+C_+F^{\rm c}_z/2)}
\end{pmatrix},
\end{align}
where
\begin{align}
\lambda' \equiv& \sqrt{\left(\tp+C_+\frac{F_z^{\rm c}}{2}\right)^2+\left(\frac{q}{2}+C_- d\right)^2},\\
 \theta'\equiv&{\rm arg}\left(\frac{q}{2}+C_-d\right).
\end{align}
Since $n^{\rm nc}_{\pm 1}$ and $d^{\rm nc}$ are self-consistently determined so as to satisfy Eq.~\eqref{eq:tnnc_Li},
we obtain the relation between $F_z^{\rm nc}$ and $\dnc$ as
\begin{align}
 \frac{\dnc}{F_z^{\rm nc}} 
&= \frac{\displaystyle\sum_{\nu=\pm}\bar{u}_{-1}^{(\nu)*}\bar{u}_1^{(\nu)}
\textrm{Li}_{3/2}(e^{-\beta\kappa_\nu})}
{\displaystyle\sum_{\nu=\pm}\left[\bar{u}_1^{(\nu)*}\bar{u}_1^{(\nu)}-\bar{u}_{-1}^{(\nu)*}\bar{u}_{-1}^{(\nu)}\right]
\textrm{Li}_{3/2}(e^{-\beta\kappa_\nu})}
 \nonumber\\
&=  \frac{q/2+C_-d}{(2C_+-C_-)F_z}.
\label{eq:dnc_BA}
\end{align}

Equations~\eqref{eq:dc_BA} and \eqref{eq:dnc_BA} are rewritten as a linear equation of $F^{\rm c}_z$ and $F^{\rm nc}_z$:
\begin{align}
 \begin{pmatrix} 
  2c_1 \dc -C_- \dnc - q/2 & (2C_+-C_-)\dc \\
  -(2C_+-C_-)\dnc & C_- \dc - 4c_1\dnc+q/2
 \end{pmatrix}
 \begin{pmatrix} F^{\rm c}_z \\ F^{\rm nc}_z \end{pmatrix} =0.
\end{align}
From the condition that $F^{\rm c}_z$ and $F^{\rm nc}_z$ have a non-trivial solution, 
we obtain 
\begin{align}
 q &\cong 2(C_+-C_-)\dc\left( 1 + \frac{4C_+-C_-}{C_+}\frac{\dnc}{\dc}\right),\\
&= 4c_1 \dc \left(1+\frac{3c_0+5c_1}{c_0+c_1}\frac{\dnc}{\dc}\right),
\end{align}
where we have expanded $q$ to first order in  the parameter $\dnc/\dc$.
Since $\dc=-n^{\rm c}/2$ at the BA-P boundary, we obtain the boundary $q_{\rm b}$ as Eq.~\eqref{eq:BA-P_boundary}.

In the laboratory frame, $\nnc$ is related to that in the rotated frame as
\begin{align}
{\bm n}^{\rm nc(lab)} &= e^{-i f_y\pi/2} \nnc e^{if_y\pi/2},
\label{eq:frame}
\end{align}
from which $\dnc$ is rewritten in terms of $\nnc$ in the laboratory frame as
\begin{align}
 \dnc \equiv n^{\rm nc}_{-1,1}= \frac{1}{2}\left( n^{\rm nc(lab)}_{1,1}+n^{\rm nc(lab)}_{-1,1}-n^{\rm nc(lab)}_{0,0}\right).
\end{align}

\subsection{Neglecting the off-diagonal part}\label{FHF2}
When we neglect the off-diagonal part of ${\bm n}^{\rm nc}$ in the laboratory frame,
\begin{align}
{\bm n}^{\rm nc(lab)} = \begin{pmatrix}
n_1^{\rm nc(lab)} & 0 & 0 \\
0 & n_0^{\rm nc(lab)} & 0 \\
0 & 0 & n_{-1}^{\rm nc(lab)}
\end{pmatrix},
\end{align}
the non-condensed component has no transverse magnetization,
which means $F_z^{\rm nc}=0$, i.e., $n^{\rm nc}_1=n^{\rm nc}_{-1}$, in the rotated frame.
Hence, the calculation for the condensate part is the same as that for the full-HF calculation if we impose $n^{\rm nc}_1=n^{\rm nc}_{-1}$.
Equation~\eqref{eq:dc_BA} then reduces to
\begin{align}
 \frac{\dc}{F_z^{\rm c}} &= \frac{q/2+C_-\dnc}{(C_+-C_-)F^{\rm c}_z},
\end{align}
from which we obtain Eq.~\eqref{eq:BA-P_boundary_diagonal}.


%

\end{document}